\documentclass[12pt]{iopart}
\usepackage{graphicx}
\usepackage{iopams}
\begin{document}
\title{Mode-coupling theory predictions for a limited 
valency attractive square-well model 
}
\author{E. Zaccarelli$^{1,2}$ 
\footnote{E-mail: emanuela.zaccarelli@phys.uniroma1.it},
I. Saika-Voivod$^{3}$, A. J. Moreno$^{4}$, E. La Nave$^{1,2}$,\\ 
S. V. Buldyrev$^{5}$,
F. Sciortino$^{1}$ and P. Tartaglia$^{6}$}
 \address{
$^1$ Dipartimento di Fisica and CNR-INFM-SOFT, Universit\`a di Roma `La Sapienza', P.le A. Moro~2, I-00185, Roma, Italy,\\
$^2$ ISC-CNR, Via dei Taurini 19, I-00185, Roma, Italy,\\
$^3$ Department of Chemistry, University of Saskatchewan,
Saskatoon, Saskatchewan, S7N 5C9, Canada,\\
$^4$ Donostia International Physics Center,
Paseo Manuel de Lardizabal 4,
E-20018 San Sebasti\'an, Spain,\\
$^5$ Yeshiva University,  Department of
Physics, 500 W 185th Street New York, NY 10033, USA,\\
$^6$  Dipartimento di Fisica and CNR-INFM-SMC, Universit\`a di Roma `La Sapienza', P.le A. Moro~2, I-00185, Roma, Italy.}

\begin{abstract}
Recently we have studied, using numerical simulations, a  limited
valency model, i.e. an  attractive square well model with a constraint on the maximum number of  bonded neighbors.  
Studying a large region of  temperatures $T$ and
packing fractions $\phi$, we have estimated  the location of the liquid-gas phase separation spinodal and the loci of dynamic arrest, where the system
is trapped in a disordered non-ergodic state.
Two distinct arrest lines for the system are present in the system: a {\it (repulsive) glass} line at high packing fraction, and a {\it gel} line at low $\phi$ and $T$. The former is essentially vertical ($\phi$-controlled), while the latter is rather horizontal ($T$-controlled) in the $(\phi-T)$ plane.  
We here complement the molecular dynamics results with mode coupling
theory calculations, using the numerical  structure factors as input.  We find that the theory predicts a repulsive glass line  --- in satisfactory agreement with
the simulation results --- and an attractive glass line
which appears to be unrelated to the gel line.
\end{abstract}


\section{Introduction}
In recent years, investigation of structural arrest in colloidal systems has witnessed a renewed and large interest in the scientific community since new
phenomena have been identified, in particular when the particles
interact via an attractive potential of range short enough compared to the size of the colloid~\cite{Sci05a}.
Theoretical~\cite{Fab99a,Ber99a,Daw00a},
numerical~\cite{Pue02a,Fof02a,Zac02a} and
experimental~\cite{Mal00a,Pha02a,Eck02a,Chen03a,Pon03a,Gra04a} studies
have shown the existence of two different mechanisms responsible for the slowing down characteristic of structural arrest. At high
temperature  $T$  and high packing fractions $\phi$ of the dispersed phase caging effects prevail and produce the typical repulsive glass behavior. At low  $T$, and slightly smaller $\phi$, another mechanism sets in, due to the stickiness of the particles, and generates a so-called attractive glass~\cite{Sci02a}. It was hypothesized at the beginning that the latter mechanism could be one of the routes to the formation of a gel at low densities~\cite{Ber99a}. Many
investigations of these phenomena have been made~\cite{Man05a,Lu06a}, but most of them
were faced with difficulties related to the existence of a two-phase
region at low $\phi$. In fact it was unambiguously
shown, for the case of short range attractive potentials, that the arrest line intersects the binodal line at its high
volume fraction side~\cite{Zaccapri,Fof05a}. The crossing may be avoided if a lowering and
shrinking of the two-phase coexistence region is achieved.  

One possibility to limit or suppress phase separation is to consider
the effects of a long-range repulsion complementing the short-range
attraction. This  type of potential, able to mimic effects of screened electrostatic interactions, properly describes interactions in charged  
colloidal suspensions. This route has been actively pursued very recently in experiments~\cite{Strad04,Bagl04,Bartlett04},
simulations~\cite{Sci04a,Coni04,Imp04,Mos04a,Sciobartlett} and
theory~\cite{ChenPRE,Liu05,Wu05a,Tarzia05}.

Another possibility to limit or suppress phase separation, which does
not invoke the presence of repulsive long range repulsion is offered
by saturation of bonding, i.e. by a limit on the maximum number of
bonded interactions. To prove this mechanism, we have devised a model
where an ad-hoc constraint is adopted, in addition to a square well
(SW) interaction.  We studied numerically a saturated SW
model first introduced by Speedy and Debenedetti~\cite{Spe94,Spe96,notaspeedy},
also named limited valency model.  In this model, the square well
attraction between two particles is constrained to a maximum number of
bonds $n_{\rm max}$ that particles can form.  By reducing $n_{\rm
max}$ down to low values such as $3,4,5$, we found that the location
of phase separation is progressively shrunk both in $\phi$ and $T$
~\cite{Zaccagel}.  Hence, reduction in the number of bonds may avoid
the crossing between the dynamic arrest and the binodal lines, or to
move the crossing to much lower $\phi$ and $T$.  Simulating this
model, we showed that indeed the avoidance of phase separation allows
the emergence of arrested states at low $\phi$, which have quite
different physical features than glasses, both of attractive and
repulsive type. Completing the study up to very large densities
in~\cite{ZacJCP}, we detected the presence of two distinct arrest
lines: a gel line, practically flat in $T$ , governed by Arrhenius
dynamics, and a glass line, practically constant in $\phi$,
corresponding to the standard hard sphere glass transition.  All along
the gel line, the dynamic features are not reducible to those of
attractive glasses, suggesting that the two states in attractive
systems do not necessarily correspond to the same phenomenon,
and a simple extension of the attractive glass line may
not be always appropriate.

The purpose of the present work is to use the MCT in order to check the conclusions obtained through numerical simulations, to clarify in
particular the possibility of describing the low $\phi$ equilibrium gel formation. To this aim, we compare here our numerical results of Refs.~\cite{Zaccagel,ZacJCP} with predictions of Mode Coupling Theory (MCT) for the same model, using as input of the theory the
"exact" structure factors for the $n_{max}$ model,
calculated numerically. 

\section{Overview of the simulation results}

\begin{figure}
\begin{center}
\includegraphics[width=11cm,angle=0.,clip]{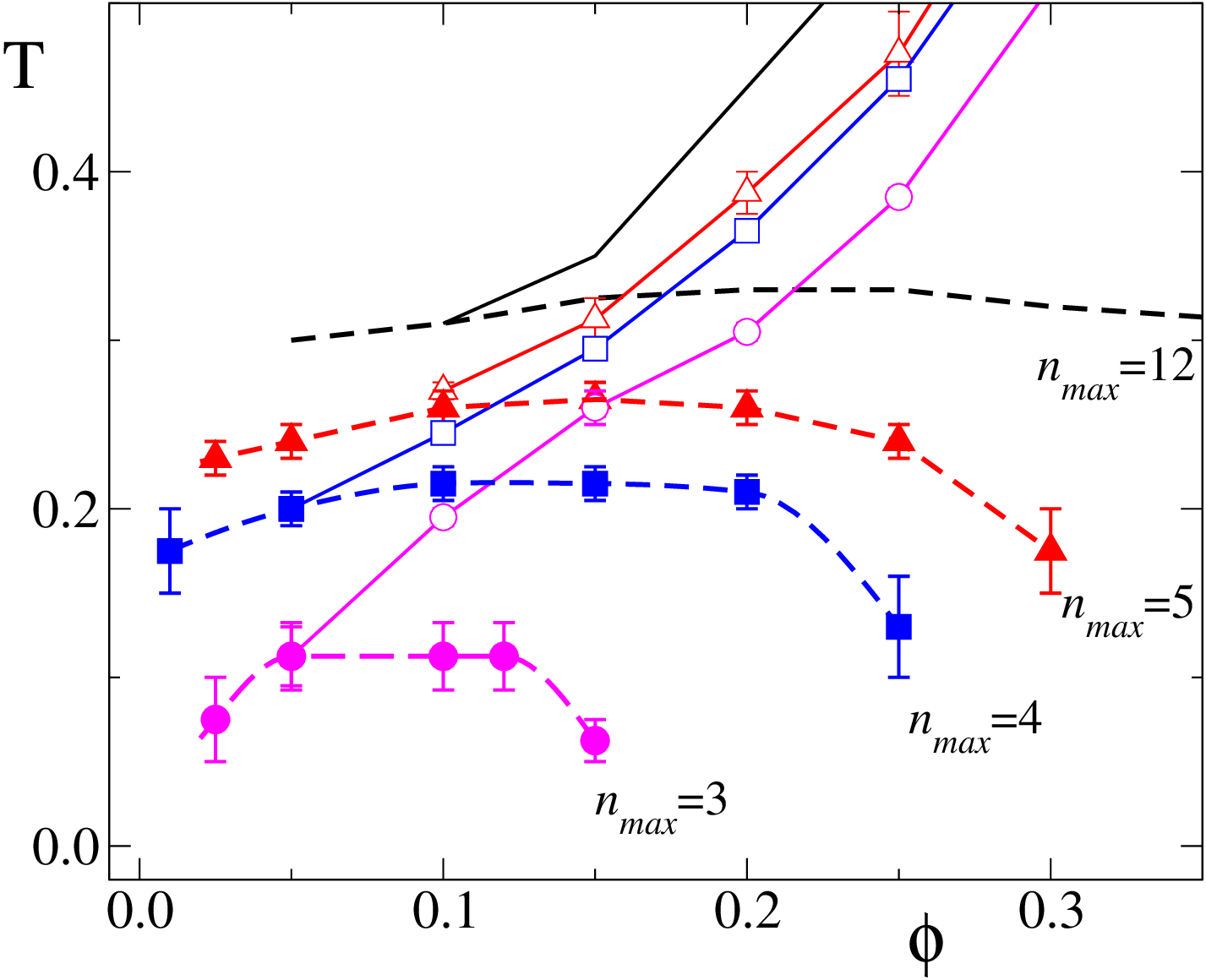}
\end{center}
\caption{Spinodal and percolation lines (dashed and solid lines respectively) of the limited valency model for $n_{\rm max} =3 $ (circles), $n_{\rm max} = 4$ (squares), $n_{\rm max} = 5$ (triangles) and $n_{\rm max} = 12$ (no symbols).}
\label{fig:new-phase-diagram}
\end{figure}

The limited valency model described above has been recently studied in detail
and we summarize here the main results of the numerical
investigation~\cite{Zaccagel,ZacJCP}. Molecular dynamics was performed in a  square-well attractive system of 
$N=10^4$ particles of unit mass and hard core $\sigma=1$  with a range of the well $\Delta$
such that   $(\sigma+\Delta)/\sigma=1.03$  and depth $u_0$. 
Temperature is measured in units of $u_0$, with the Boltzmann
constant $k_B=1$.  The maximum number of bonds was limited to $n_{\rm max} = 3,4,5$ as compared to the unconstrained SW, where $n_{\rm max} = 12$. Indeed, for $n_{\rm max} >6$, no significant difference in the phase behaviour was observed with respect to the standard square well model. The main findings are reported in the following list summarizing our results.
\begin{enumerate}
\item
The coexistence region in the $(\phi,T)$ plane reduces in size as
$n_{\rm max}$ decreases, the values of $\phi$ above which the spinodal  disappears are $\phi \approx 0.20$ and $\phi \approx 0.30$ for $n_{\rm
max} = 3$ and 4 respectively. The loci of percolation also shift to
lower $T$ values, and cross the two-phase region on the low $\phi$
side, as shown in Fig.~\ref{fig:new-phase-diagram}.
\item
The bonds formed due to the attraction show a lifetime which follows
an Arrhenius behavior in  $T$ .
\item
The mean squared displacement develops a plateau at low  $T$
which defines a localization length of the order of the particle
size, much larger both than the length typical of a repulsive glass
due to caging (of the order of $0.1 \sigma$) and than that of an
attractive glass (of the order of the well width).
The diffusivity  shows a power-law behaviour  in $\phi$ along constant temperature paths,
while it displays
an Arrhenius dependence on $T$ for constant volume fraction.
\item
The normalized intermediate scattering functions show a corresponding
plateau at low $T$ only for values of the momentum transfer $q$
smaller than the value corresponding to the first peak of the
structure factor. This feature is rather different from the standard
behavior close to a glass transition, and is similar to the behavior
of a chemical gel~\cite{Voi04a}.
\item
The height of the plateau for the normalized intermediate scattering
functions, i.e. the non-ergodicity factor $f_q$, has a different shape
in $q$ for the gel than both its attractive and repulsive glass
counterparts.
\end{enumerate}
Based on these facts we concluded that the system shows a {\it glass}
line at high $\phi$ and a {\it gel} line at low  $T$ 
and low $\phi$.  More important, the dynamical behavior observed close to the gel line  does not seem to 
allow the identification of the gel with the attractive MCT glass.
  
\section{Mode Coupling Theory}
We solve the MCT equations  to locate the ideal glass line(s) in the $(\phi-T)$ plane.  The MCT, starting only from the structural information contained in the static structure factor $S(q)$, provides indication of the onset of non-ergodic behavior~\cite{goetze}.  Although the theory is not strictly based on the hypothesis of pair additivity for the
interaction potential, its main (uncontrolled) approximation is the
factorization of higher order correlation functions into products of  pair
correlation functions~\cite{goetze,Zac01a}. Since the $n_{\rm max}$
model incorporates many-body terms in the Hamiltonian\cite{Zaccagel,ZacJCP}, we cannot
expect the theory to work at its best. Perhaps higher order
correlations should also be considered, as in the case of silica, a
tetrahedral network-forming liquid, where the triplet correlation
function $c_3$ provides a relevant contribution to the MCT
kernel~\cite{Sci01aPRL}. Here, we limit ourselves to the standard
version of MCT calculation, because the random nature of the $n_{\rm
max}$ bonds along the particle surface does not produce any angular
constraint, retaining the full sphericity for the model.
However, it would be interesting to check in a future work whether the inclusion of  
the corresponding 
three-point correlators  calculated within MD simulations produce any significant difference
in the MCT results.

MCT is able to predict the full time evolution for the density-density autocorrelation function  $\Phi(q,t)$ as a function of the
momentum transfer $q$ and $t$, through coupled non-linear
integro-differential equations.  For a given interaction potential,
through the knowledge of $S(q)$, the memory kernel entering in the
nonlinear term of the MCT equations can be evaluated and the equation
solved for various values of $q$.  In particular, the non-ergodicity
transition leading to structural arrest is obtained by performing the
limit $t\rightarrow \infty$ in the equations. Defining the
non-ergodicity factor as the long-time limit of $\Phi$ correlator
$\lim_{t \rightarrow \infty} \Phi(q,t) = f(q)$, $f(q)$ is found to be
the solution of~\cite{goetze},
\begin{equation}
\label{eq-nonergodicity}
{{f(q)} \over {1-f(q)}} = m(q)
\end{equation}
where the memory kernel $m(q)$ is quadratic in the correlator itself,
\begin{equation}
\label{eq-kernel}
m(q) = {{1}\over{2}}\int {{d^{3}k}\over{(2\pi)^{3}}}{\cal V}( {\bf
q},{\bf k})f({k})f(|{\bf q}-{\bf k}|).
\end{equation}
The vertex functions ${\cal V}$, the coupling constants of the theory,
are
\begin{eqnarray}
\label{eq-vertex}
{\cal V}( {\bf q},{\bf k}) = \frac{\rho}{ q^{4}}\left[{\bf q} \cdot(
{\bf q} - {\bf k}) ~c({|{\bf q}-{\bf k}|})+ {\bf q}\cdot {\bf
k}~c({k})\right]^{2} S({q}) S({k}) S({|{\bf
q}-{\bf k}|})
\end{eqnarray}
and depend only on the Fourier transform of the direct correlation
function $c(q)$, or equivalently on $S(q)$, and on the number density
$\rho$.  
In the $A_2$ bifurcation scenario of MCT~\cite{goetze} 
the solutions of Eq.~(\ref{eq-nonergodicity}) jump from zero to a finite
value at the ideal glass transition. The locus of the fluid-glass
transition can be calculated varying the control parameters of the
system, $\phi$ and $T$.  For a square well model, there is an additional
control parameter, that is the range of attraction $\Delta$. In this
case, higher order bifurcations of the solutions arise when $\Delta <
\Delta^*\sim 0.041$~\cite{Daw00a}. In this case, two distinct glassy
solutions appear, a repulsive and an attractive glass respectively,
with different non-ergodicity parameters and mechanical
properties~\cite{Zaccamechanical}.
\begin{figure}
\begin{center}
\includegraphics[width=11cm,angle=0.,clip]{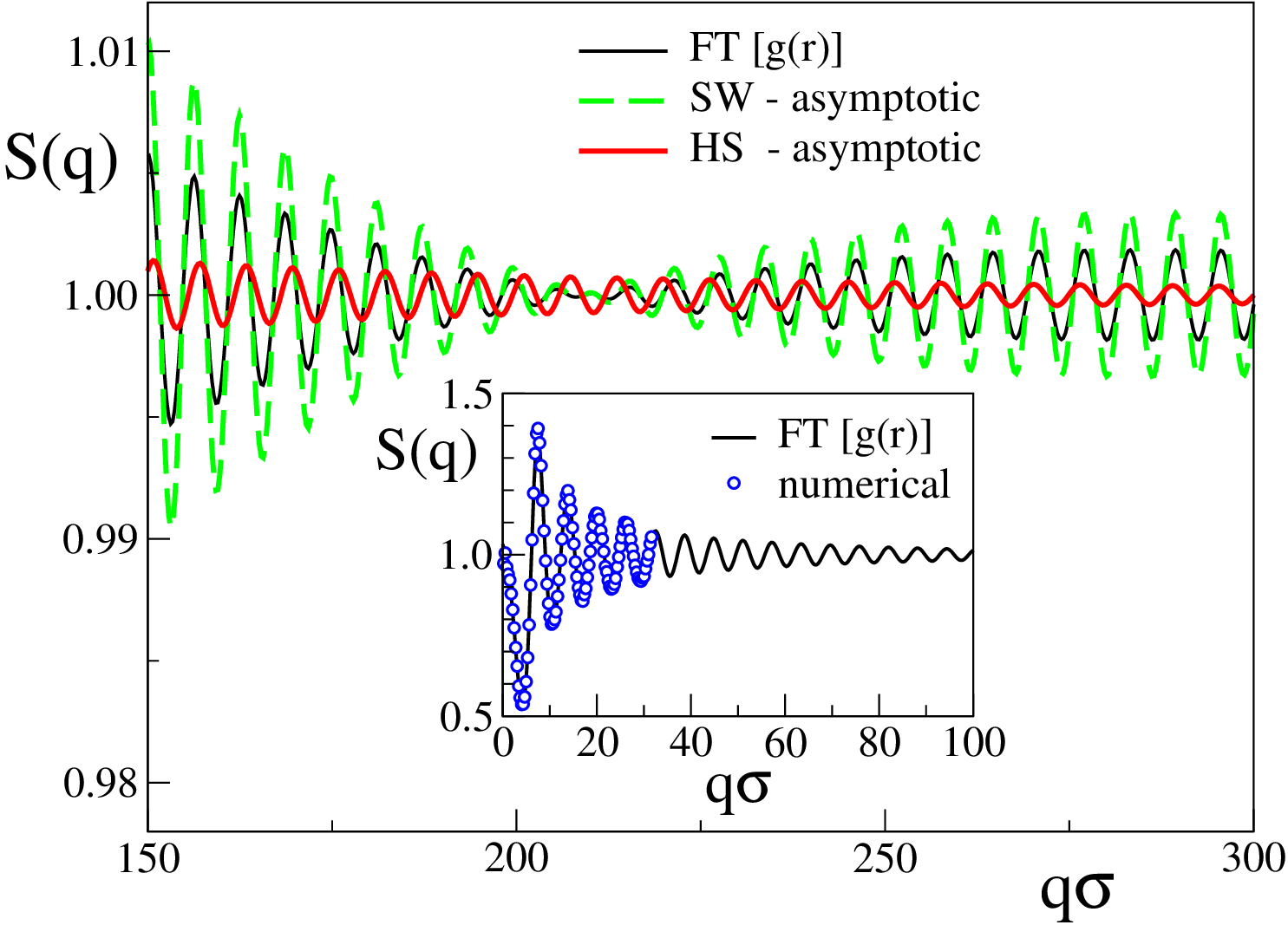}
\end{center}
\caption{Inset: Static structure factor for $\phi=0.20$ and $T=0.2$ used as input in MCT
calculated directly from simulations (points), and by Fourier transform
of the $g(r)$ (line). Main: Enlargement of the large-$q$ behavior
displaying `beats'. For comparison we also report the corresponding
asymptotic SW (differing only by a factor $A$ in amplitude) and HS
(out of phase) results.}
\label{fig:sq-MCT}
\end{figure}

We calculate the MCT ideal glass transition line for $n_{\rm max}=3$
using as input the numerical $S(q)$, "exact" within numerical
precision.  In the evaluation of the MCT kernel of
Eqn.~(\ref{eq-kernel}), it is crucial to integrate over all $q$
contributing to the memory function. For short-range attractive
potentials, it is important to integrate up to very large $q$ values,
since the information of the potential shape is coded into the large
$q$ region.  From a numerical point of view, it is convenient to
evaluate $S(q)$ at large $q$ by Fourier transforming the pair
distribution function $g(r)$ and at small $q$, by direct evaluation of
$S(q)$ in $q$-space.  Indeed, bonding in $g(r)$ is reflected by a very
large and constant signal between $\sigma$ and $\sigma+\Delta$, which
fully accounts for the large $q$ behavior of $S(q)$.  For the same
reason, the large-$q$ `beats'~\cite{Zaccakonstanz} in $S(q)$ have the
same shape as the standard SW and carry information of the short-range
potential.  The asymptotic form of $S(q)$ for large-$q$, both for the
SW and the $n_{\rm max}$ model, is
\begin{eqnarray}
\!\!\!\!\!\!
S_{asympt}(q)\!\!\!\ &=&\ \!\!1-\frac{A}{q^3}
\left\{\sin{(q\sigma)}-q\sigma \cos{(q\sigma)}\right.\nonumber\\&+& 
\left. 
\!\!
(e^{\beta u_0}-1)[q(\sigma+\Delta) \cos{(q(\sigma+\Delta))} 
\right.
\nonumber\\&-&
\left.\!\!
\sin{(q(\sigma+\Delta))}+\sin{(q\sigma)}-q\sigma\cos{(q\sigma)}]
\right\}\!\!.
\label{eq:asy}
\end{eqnarray}
The amplitude $A$  (which depends on $T$ and $\phi$) is different for the SW and $n_{\rm max}$ models, being smaller in the $n_{\rm max}$ case, due to the reduced number of
bonded neighbors.  Combining information from $S(q)$ and $g(r)$, an
accurate description of $S(q)$ over the entire relevant $q$ range is
obtained, as shown in Fig.~\ref{fig:sq-MCT}.  In the SW case, the
large-$q$ oscillations resulting from the narrow width of the square
well potential are largely responsible for the MCT attractive glass
transition~\cite{Daw00a}.
\begin{figure}
\begin{center}
\includegraphics[width=11cm,angle=0.,clip]{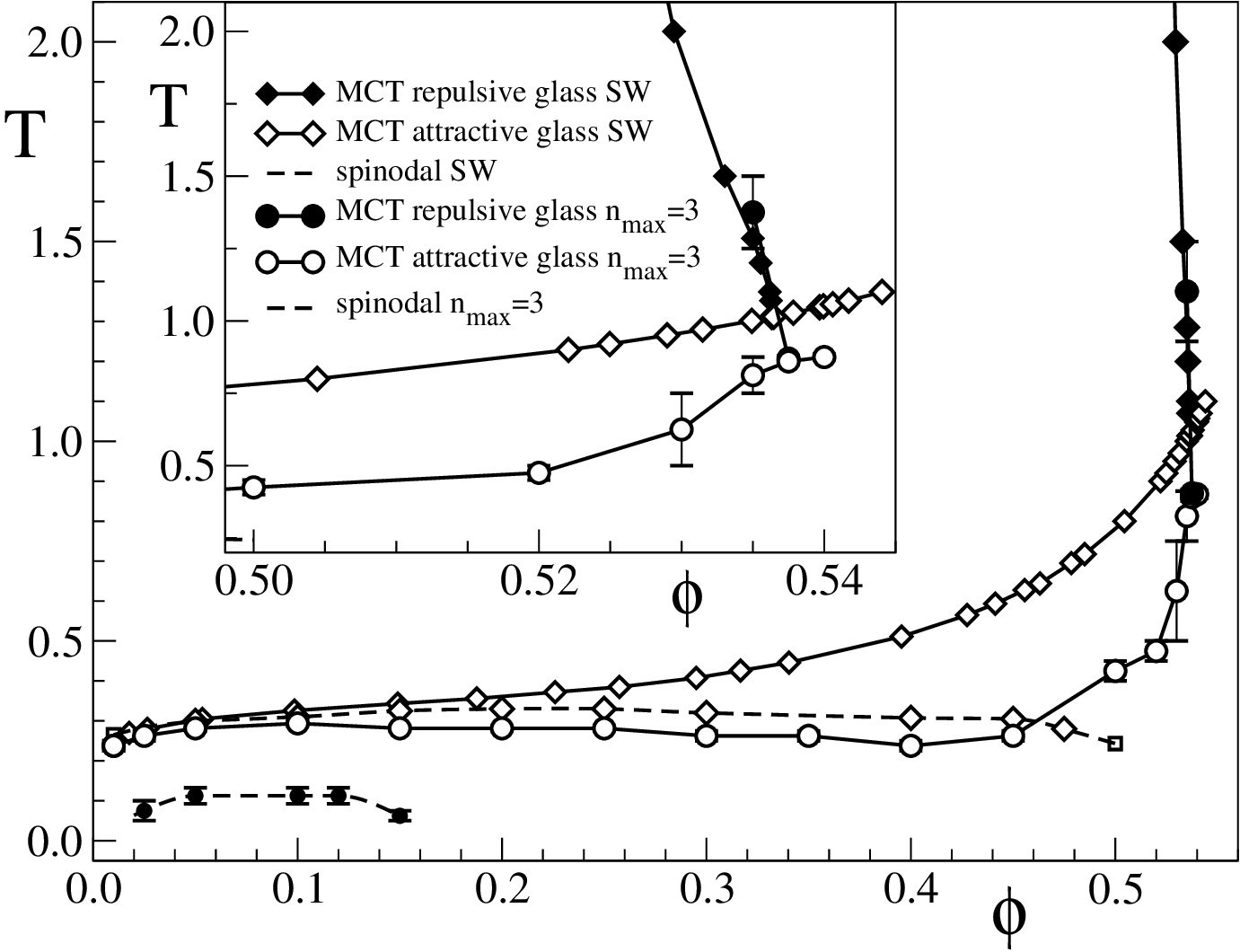}
\end{center}
\caption{ 
Comparison of MCT predictions and spinodal lines for the SW case
and for the $n_{\rm max}$ model with $n_{\rm max}=3$.  The inset shows
an enlargement of the high $\phi$ region, to visualize the attractive
and repulsive lines and the glass-glass transition. 
}
\label{fig:mct-large}
\end{figure}
We solve the MCT equations in $q$-space in the window from $0$ to
$600\sigma^{-1}$, with a mesh of about $0.3\sigma^{-1}$, for a total of $2000$
$q$-vectors.  We bracket the MCT arrest line by locating two adjacent
state points along isochores where a liquid and glassy solutions are
respectively found.  The resulting MCT predictions are reported in
Fig.~\ref{fig:mct-large}.

At high densities, MCT results are very similar to those found for the
simple SW~\cite{Daw00a,Zaccamechanical} for the $\Delta=0.03$ case.
Even in the $n_{\rm max}$ case, the MCT equations predict two distinct
glass lines, respectively an attractive and a repulsive one, and a
glass-glass transition ending in an $A_3$ singularity~\cite{sperl}.  A
comparison between the MCT predictions for the SW and for the $n_{\rm
max}$ model is reported in Fig.~\ref{fig:mct-large}. The glass lines for
both models converge to the hard-sphere result at high $T$ and to the
same locus for $\phi \rightarrow 0$.  This is consistent with the
expectation that at high $\phi$ arrest is driven by packing, while at
low $\phi$ the constraint on the maximum number of bonded neighbors
becomes irrelevant and hence the two models tend to become similar.
Differences with respect to the SW case are observed at intermediate
$\phi$.

\begin{figure}
\begin{center}
\includegraphics[width=11cm,angle=0.,clip]{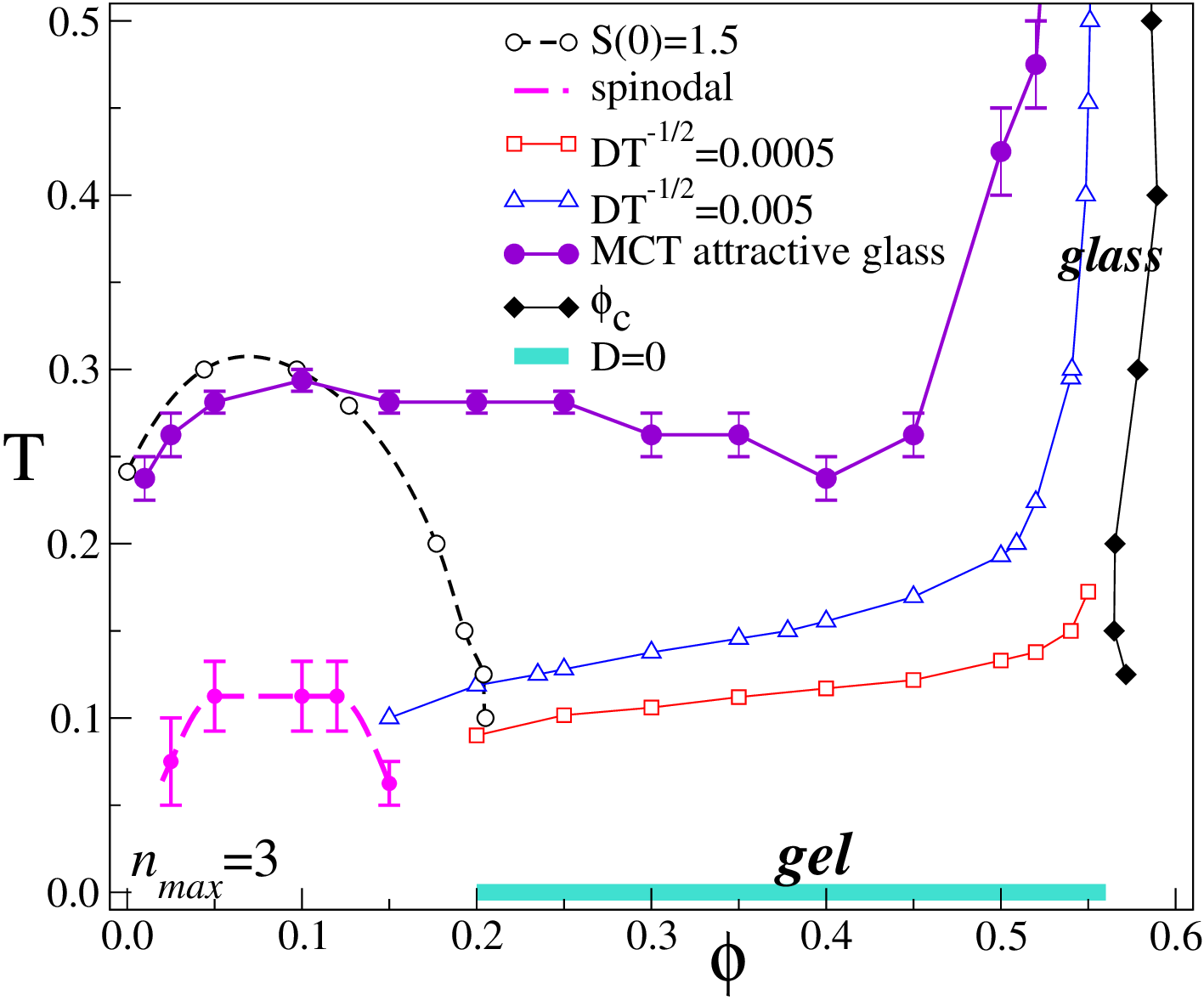}
\end{center}
\caption{Summary of the thermodynamic and kinetic phase diagram for 
$n_{\rm max}=3$,
including spinodal (dashed lines with filled circles), iso-diffusivity
loci where $D/\sqrt{T}= 0.005, 0.0005$ (lines with triangles and
squares), iso-$S(0)$ locus where $S(0)=1.5$, MCT lines (attractive
and repulsive), extrapolated {\it glass}, labeled 
as $\phi_c$, and
{\it gel}, labeled as $D=0$, lines respectively 
from power law and Arrhenius fits
($T_c=0$)\protect\cite{ZacJCP}. }
\label{fig:summary}
\end{figure}

In the SW case the ideal MCT attractive glass line is rather flat,
monotonically decreasing in $T$, almost merging into the spinodal on
the left side of the critical point~\cite{Zaccamechanical}.  The
$n_{\rm max}$ model shows instead a non-monotonicity of the
$T$-dependence of the attractive line with decreasing $\phi$, causing
the presence of a minimum in  $T$  for $\phi \approx 0.4$ and
$T\approx 0.24$.  The significant suppression of the attractive glass
line in the $n_{\rm max}$ model at intermediate $\phi$ arises from the
decrease in the number of bonded nearest neighbors as compared to the
SW case.  The attractive glass line has a maximum around $\phi \approx
0.10$ and $T\approx 0.3$, before turning down following the spinodal
as in the SW case, on the left of the critical point.  It is possible
that the shape for $\phi \lesssim 0.1$ in the MCT line is an echo of
the underlying increase of the compressibility at low $\phi$ due to
the close-by spinodal, since at low $\phi$ the MCT line roughly follows
in shape the loci of constant $S(0)$ (see Fig.~\ref{fig:summary}).
\begin{figure}
\begin{center}
\includegraphics[width=11cm,angle=0.,clip]{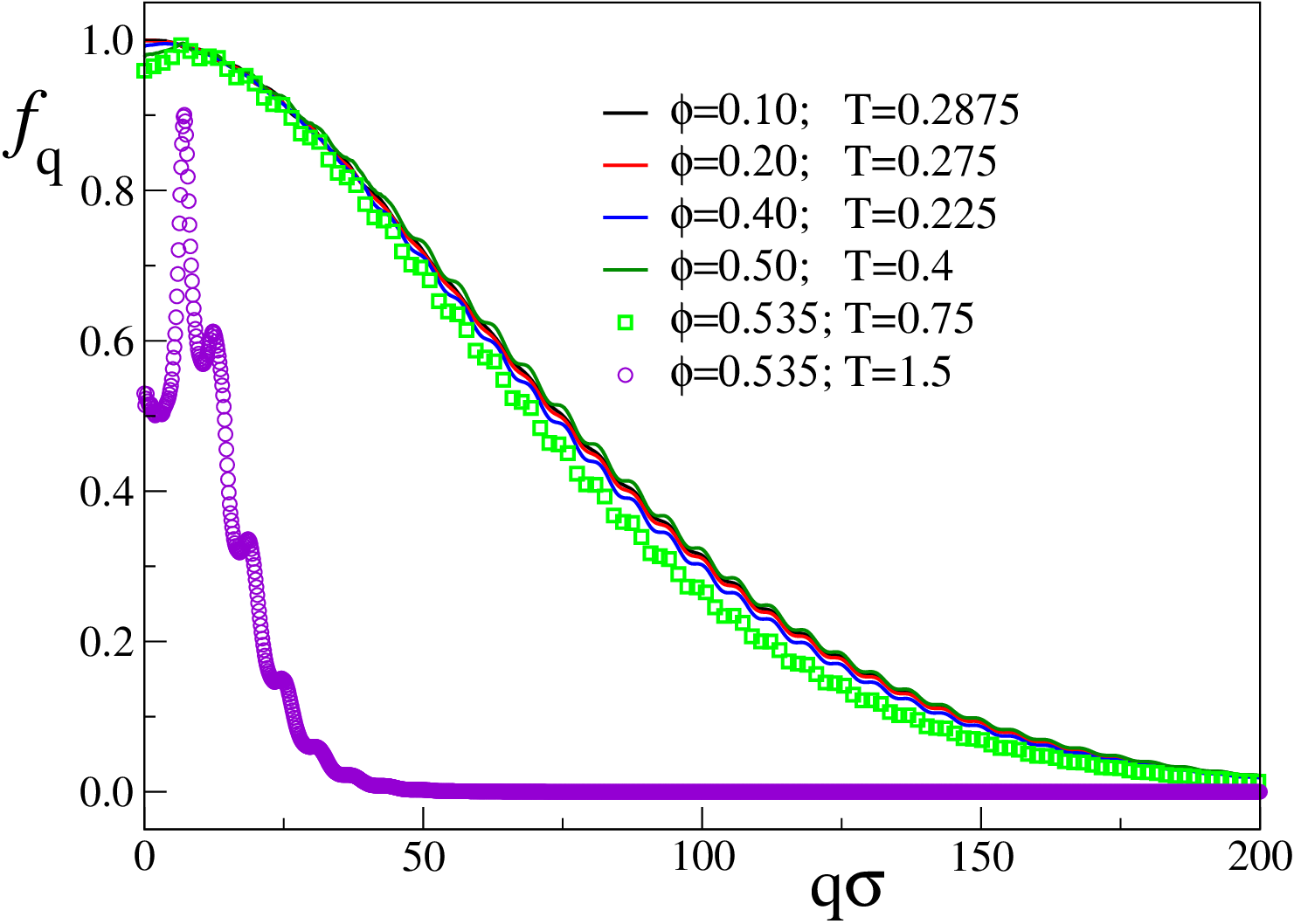}
\end{center}
\caption{ 
Critical non-ergodicity
parameters $f_q$ calculated within MCT. Note that at low $\phi$ along
the attractive glass line, $f_q$ are indistinguishable. }
\label{fig:fq-MCT}
\end{figure}
\begin{figure}
\begin{center}
\includegraphics[width=11cm,angle=0.,clip]{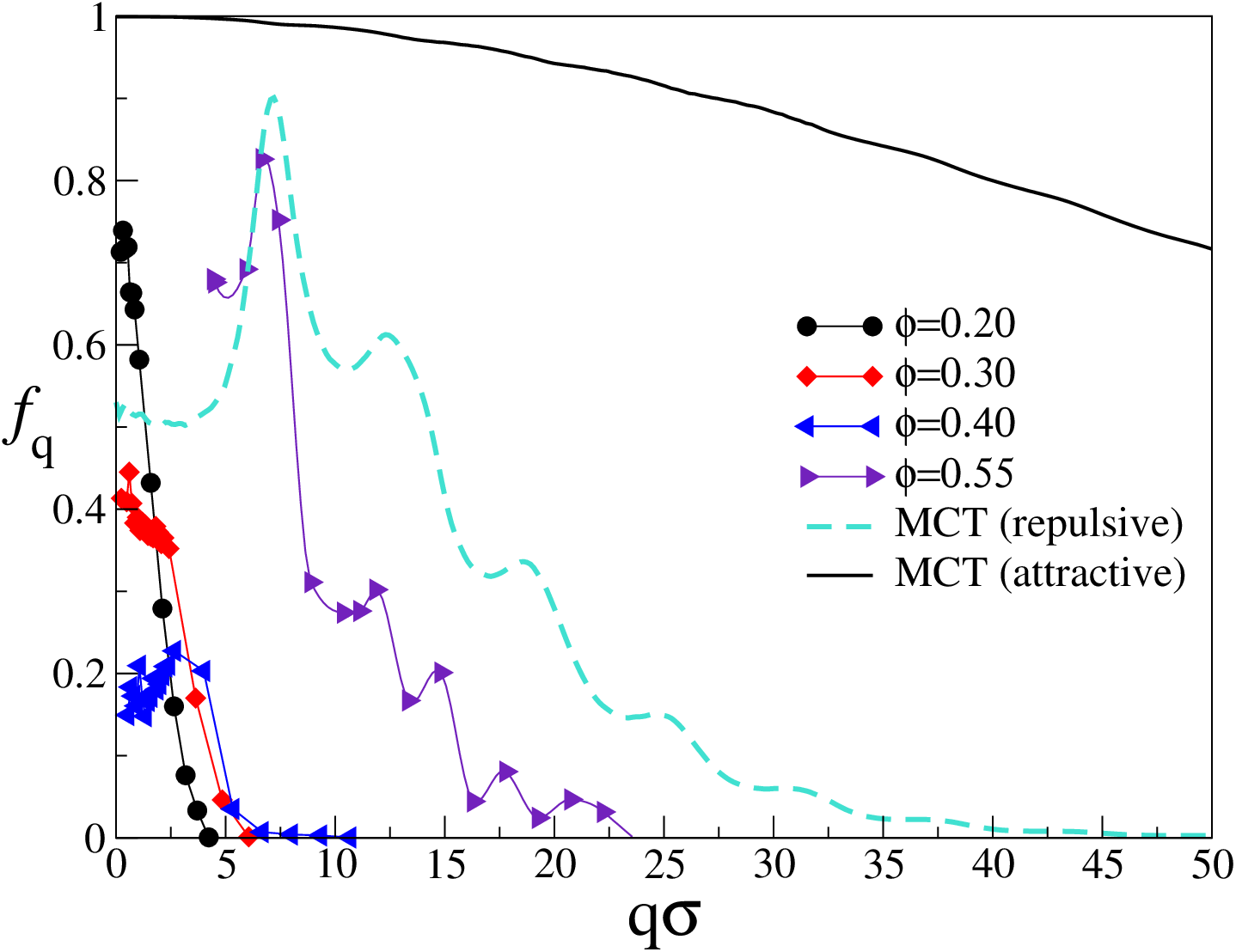}
\end{center}
\caption{Critical non-ergodicity parameter $f_q$ calculated from the
simulations (i.e. along the iso-diffusivity line $DT^{-1/2}=0.0005$), via a stretched exponential fit of the $\phi$
correlators. Note that $f_q$ never resembles that of an
attractive glass. Lines are guide to the eye.}
\label{fig:fq-simu}
\end{figure}

In the SW, the attractive and the repulsive glass lines differ
essentially in the $q$-dependence of the non ergodicity parameter
$f_q$.  The theory also allows us to calculate $f_q$ and its critical
value at the glass transition.  As compared to the repulsive glass,
the attractive glass is characterized by much larger $f_q$ values,
extending to much larger $q$.  These features are also displayed in the
theoretical calculations for the $n_{\rm max}$ model. Along the
attractive glass line, for all $\phi$ values, from about $\phi=0.05$ up to $\phi=0.54$, the critical 
non-ergodicity parameter, i.e. $f_q$ at the MCT transition, does not change
significantly. Fig.~\ref{fig:fq-MCT} shows the full $q$-vector
dependence for both types of glasses.  We have also verified that in
the attractive glass phase $f_q$ is significantly dependent on $T$.

For comparison, we plot in Fig.~\ref{fig:fq-simu} the non-ergodicity
parameters estimated from the simulations, via a stretched exponential
fit of the $\phi$ correlators~\cite{ZacJCP}, along the iso-diffusivity line $DT^{-1/2}=0.0005$.  A transition from a gel to a glass is evident. However, 
the shape of $f_q$ never  resembles that of an attractive glass.  The gel is characterized by non-ergodic features only at large length-scales, due
to the mobility of the network in the available free space, while an
attractive glass is strictly confined within the bond length.  The
main difference between theoretical and simulation results is that,
although also in the gel bonds are on average permanent on the time scale of
 the simulations at low $T$~\cite{ZacJCP}, the limited
number of neighbours accounts for residual motions of the particles. Hence,
particles are confined by the attractive well width only relative
to each other, but still can freely explore up to their diameter
length without ever breaking the network. 
In the theoretical
calculations, however, the presence of the short-range bonds,
manifested in the large-$q$ tail of $S(q)$, is responsible for the
non-ergodic transition.

Already at the SW level, the theory strongly overestimates the
tendency to form an attractive glass~\cite{Zac03a,Zac04b}. Compared with
numerical simulation~\cite{Sci03a}, the theoretical attractive glass
line has to be shifted approximatively by a factor of 3 to 4 in $T$,
while the repulsive glass line needs only a 10 per cent adjustment in
$\phi$.  The correction of the location of the ideal MCT lines
provokes a major effect: the attractive glass line does not lie above
the critical point but meets the spinodal at low $T$ on the right
side~\cite{Zaccapri}.  If the overestimate of the attractive glass line
is properly taken into account, then one has to conclude that it is
not possible to form arrested states at low $\phi$ without the
intervention of a phase separation. Recent simulation studies confirm
that this is the case independently from the width of the attraction
range~\cite{Fof05a}. It is also necessary to recall that the ideal MCT
glass lines, especially when energetic caging is dominant, have to be
interpreted as cross-over lines, from a power-law to an activated
dependence of dynamic properties~\cite{attivatiinMCT}.  
We can summarize the dynamical arrest behavior in
Fig.~\ref{fig:summary}.  One locus of arrest is found at high $\phi$,
rather vertical and corresponding to the hard-sphere glass transition.
This locus is quite well described by MCT.  Very different is the
situation concerning the low $T$ slowing down.  The isodiffusivity
lines suggest a rather flat arrest line. Two different loci could be
associated with arrest at low $T$. One defined by the $T_c$ of the
power-law fits of the diffusivity and one at $T=0$ associated with the
vanishing of $D$ according to the Arrhenius law~\cite{ZacJCP}. It
would be tempting to associate the $T_c$-line with the attractive
glass line predicted by MCT and interpret the wide region between the
two lines as a region of activated bond-breaking
processes~\cite{Zac03a}.

Results for the present model provide evidence that a gel line can not
necessarily be considered an extension of the attractive glass
line. Here, a gel never transforms into an attractive glass, but only
into a repulsive glass with increasing $\phi$. The competition between
the two arrested states seems also to produce anomalies in the
dynamics of the same kind as those found in the presence of MCT higher
order singularities~\cite{ZacJCP}.  In particular, these anomalies
correspond to a logarithmic relaxation for the density correlators in
the liquid region close to the singularity, accompanied by a
sub-diffusive behaviour for the mean squared
displacement\cite{Sci03a}.
\section{Conclusions}
The aim of this work was to compare results for the dynamical arrest
in the $n_{\rm max}$ model for $n_{\rm max}=3$ with MCT predictions.
Using the numeric $S(q)$, "exact" within the precision of the statistical averages, we have solved the MCT equations and
evaluated the glass lines.  The theoretical results differ only
slightly from the SW case and do not provide any indication of a gel
line. According to the theory an attractive glass line (i.e. with the
typical features of the attractive glass as localization length
 $\sim \Delta^2$ and
very wide $f_q$) should be present. From a theoretical point of view,
such a line arises from the large $q$ oscillations in $S(q)$,
i.e. from the $q$-space signature of the short-range bonding.
Hence, it has the
same origin as in the SW.  In the simulation of the $n_{\rm max}$
model the bond localization is not observed, neither in the MSD nor in
the width of $f_q$. We believe this is due to the fact that, although
bonding is present, particles are confined by the potential well only
relative to each other.  Indeed, at low $\phi$, oscillations of parts
of the connected network are possible, preventing the observation of
the MCT mechanism for arrest.  At small $\phi$, even at extremely low
 $T$, when the bond lifetime becomes comparable to the
simulation time and bonds between particles are essentially permanent,
it was shown in~\cite{ZacJCP} that the plateau of the MSD remains of
the order of the particle size.  Of course, the inclusion in the MCT
calculation of higher order correlation functions could be important
in the study of a many-body interaction and should be
considered. However, due to the sphericity of the model, our opinion
is that this should not make a significant change, i.e. the prediction
for the attractive glass transition should be robust in this respect.

In summary, the present study provides a clear indication that 
in the present model, where the
liquid-gas phase separation can be avoided and arrest at low
$\phi$ can be explored in equilibrium conditions, the observed
arrested state is substantially different from the the low-$\phi$ extension of the attractive glass.    In the present model, in which
bonding interactions are not strongly directionally constrained,
large amplitude motions are possible within the fully bonded network. These modes significantly affect the shape of the correlation functions
and make it impossible to observe the short-range localization
characteristic of the attractive glass. 

The comparison with the solution of the MCT equations for the
$n_{\rm max}$ model (a possibility offered by the spherical symmetry
of the interaction potential) confirms that MCT significantly
overestimates the role of the bonding, predicting an attractive glass
even in the present case.  The present results strongly suggest that
the attractive glass is an arrested state of matter which can be
observed in short-range attractive potentials only at relatively high
$\phi$, being limited by the spinodal curve.  When the inter-particle
potential favors a limited valency, arrest at low $\phi$ in the absence of phase separation becomes
possible but with a mechanism based on the connectivity properties of
the particle network.  The dynamic features of this slowing down, 
at least in this model of  geometrically uncorrelated bonds, is
 clearly different from what would be the extension of the (attractive) glass line.

\section{Acknowledgments}
F.~S. and P.~T. acknowledge with pleasure the collaboration with
Sow-Hsin Chen in the course of various years, and dedicate this work
to the celebration of his seventieth birthday.  We acknowledge
support from MIUR-Cofin, MIUR-Firb and MRTN-CT-2003-504712.
I.~S.-V. acknowledges NSERC (Canada) for funding.  We thank W. Kob for
useful discussions.

\vskip 1cm
\section{References}
\bibliographystyle{./iopart-num}
\bibliography{./articoli,./altra}

\begin{thebibliography}{10}
\bibitem{notaspeedy}
 However, differently from \protect\cite{Spe94,Spe96}, we do not impose any
  constraint on minimal bonded loops.
\end{thebibliography}
\end{document}